# Spin-orbit qubit in a semiconductor nanowire

S. Nadj-Perge[1*], S. M. Frolov[1*], E. P. A. M. Bakkers[1,2] and L. P. Kouwenhoven[1]

[1]*Kavli Institute of Nanoscience, Delft University of Technology, 2600 GA Delft, The Netherlands*

[2]*Department of Applied Physics, Eindhoven University of Technology, 5600 MB Eindhoven, The Netherlands*

*These authors have contributed equally to this work

**Motion of electrons can influence their spins through a fundamental effect called spin-orbit interaction. This interaction provides a way to electrically control spins and as such lies at the foundation of spintronics.[1] Even at the level of single electrons, spin-orbit interaction has proven promising for coherent spin rotations.[2] Here we report a spin-orbit quantum bit implemented in an InAs nanowire, where spin-orbit interaction is so strong that spin and motion can no longer be separated.[3,4] In this regime we realize fast qubit rotations and universal single qubit control using only electric fields. We enhance coherence by dynamically decoupling the qubit from the environment. Our qubits are individually addressable: they are hosted in single-electron quantum dots, each of which has a different Landé *g*-factor. The demonstration of a nanowire qubit opens ways to harness the advantages of nanowires for use in quantum computing. Nanowires can serve as one-dimensional templates for scalable qubit registers. Unique to nanowires is the possibility to easily vary the material even during wire growth.[5] Such flexibility can be used to design wires with suppressed decoherence and push semiconductor qubit fidelities towards error-correction levels. Furthermore, electrical dots can be integrated with optical dots in p-n junction nanowires.[6] The coherence times achieved here are sufficient for the conversion of an electronic qubit into a photon, the flying qubit, for long-distance quantum communication.**

Figure 1a shows a scanning electron microscope image of our nanowire device. Two electrodes, source and drain, are used to apply a voltage bias of 6 mV across the InAs nanowire. Voltages applied to five closely spaced narrow gates underneath the nanowire create a confinement potential for two electrons separated by a tunnelling barrier. The defined structure is known as a double quantum dot in the (1,1) charge configuration.[7]

Each of the two electrons represents a spin-orbit qubit (Fig. 1b). In the presence of strong spin-orbit coupling neither spin nor orbital number are separately well defined. Instead, the two qubit states are a spin-orbit doublet, ⇑ and ⇓. Similar to pure spin states, a magnetic field $B$ controls the energy splitting between spin-orbit states $E_Z = g\mu_B B$, where $g$ is the Landé $g$-factor in a quantum dot, $\mu_B$ is the Bohr magneton. The crucial difference from a spin qubit is that in a spin-orbit qubit the orbital part of the spin-orbit wavefunction is used for qubit manipulation.[2,8]

The qubit readout and initialization rely on the effect of spin blockade.[9,10] A source-drain bias induces a current of electrons passing one-by-one through the double dot. The process of electron transfer between the dots can be energetically allowed but blocked by a spin selection rule. For instance, a (1,1)-triplet state cannot go over into a (0,2)-singlet state. This stops the left electron from tunnelling onto the right dot and thereby blocks the current. In practice the double dot becomes blocked only in a parallel configuration, *i.e.* in either a (⇑,⇑) or a (⇓,⇓) state, because antiparallel states decay quickly to a non-blocked singlet state.[11,12] By simply idling in the parameter range of spin blockade the qubits will be initialized in one of the two parallel states with equal probability. We note that spin-orbit and hyperfine interactions also mediate a slower decay of parallel states into (0,2).[7,9,10] This reduces the readout fidelity to 70-80 % (see Supplementary Information section 5.1).

A microwave frequency electric field applied to gate 4 oscillates electrons inside the nanowire (Fig. 1b). This motion can induce resonant transitions between spin-orbit states via an effect called electric-dipole spin resonance (EDSR).[2,8,13-16] Such transitions are expected when the frequency of the a.c. electric field is equal to the Larmor frequency, $f_0 = g\mu_B B/h$. At resonance the spin-orbit state of the double dot rapidly changes from parallel to antiparallel. The antiparallel state does not experience spin blockade, so the left electron tunnels to the right thereby contributing to the current. Figure 1c shows the resonance as a "V" shape which maps out the Larmor frequency in the plane of microwave frequency and magnetic field.

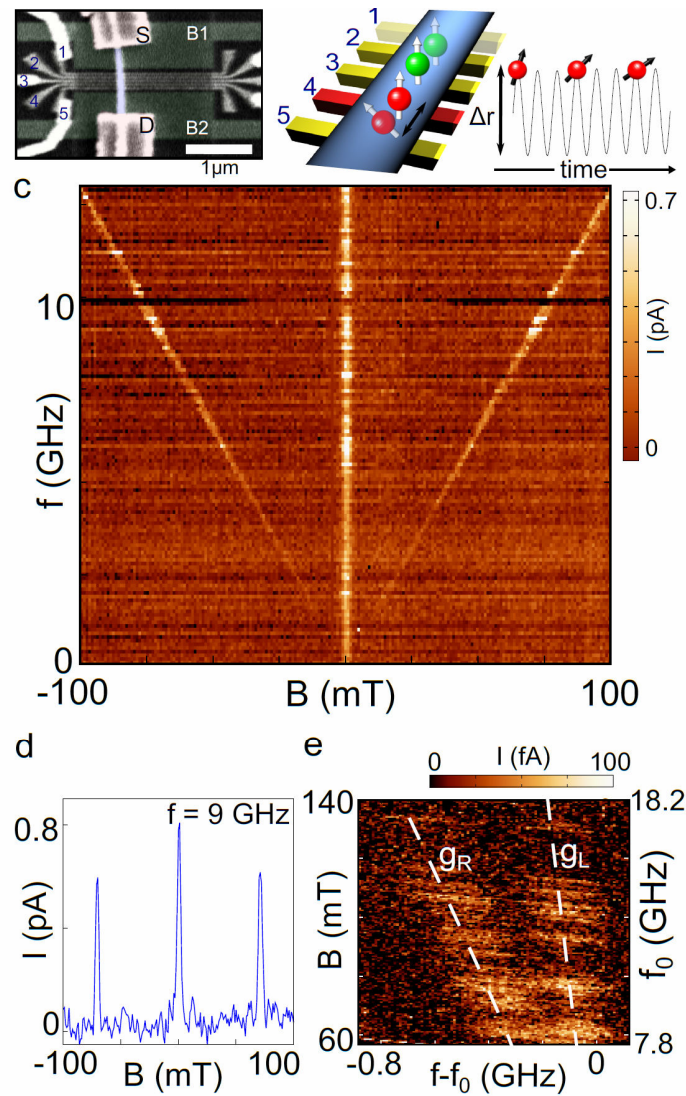

**Fig.1. Electric-dipole spin resonance. a**, scanning electron microscope image of a prototype device showing source (S) and drain (D) contacts, narrow gates 1-5 and wide gates B1 and B2. **b**, Left (red) and right (green) quantum dots are formed between gates 2-5. Microwave electric field applied to gate 4 oscillates both electrons with amplitude ~$\Delta r$ inducing EDSR. **c**, spin blockade is lifted near $B = 0$ and on resonance when $f = g\mu_B B/h$. Here microwave power $P = -42$ dBm. **d**, trace extracted from **c** at $f = 9$ GHz. **e,** Zoom in on the EDSR line which is split at high $B$ due to a difference between $g_L$ and $g_R$. At each $B$, the frequency is swept in a fixed range around $f_0 = g\mu_B B/h$ ($g = 9.28$). Current on resonance varies due to non-monotonic microwave transmission.

The "V" resonance signal vanishes in the vicinity of zero magnetic field. This behaviour is consistent with spin-orbit mediated EDSR: the effect of spin-orbit interaction must cancel at zero field due to time-reversal symmetry.[2,16] The field-dependent EDSR strength rules out a. c. magnetic field and hyperfine field gradient as possible mechanisms. A g-tensor modulation in our nanowires is estimated to be too weak to drive EDSR (see Supplementary Information section 2). The current peak near zero magnetic field arises from the hyperfine interaction between electron spin and nuclear spin bath.[11,12] From the width of this hyperfine-induced peak we extract the r.m.s. magnetic field generated by the fluctuating nuclear spins $B_N$ = 0.66±0.1 mT.[17] The width of the EDSR line at low microwave power is also consistent with broadening due to fluctuating nuclear spins (*i.e.* the side EDSR peaks and the central hyperfine peak have comparable widths in Fig. 1d).[18]

At higher magnetic field the resonance line splits up (Fig. 1e), indicating that the *g*-factors in the left and right dots, $g_L$ and $g_R$, are different. This is expected for quantum dots of different sizes since confinement changes the effective *g*-factor.[19] We measured the confinement as the orbital excitation energy at the (1,0)↔(0,1) transition and found 7.5±0.1 meV for the left dot and 9.0±0.2 meV for the right dot. A smaller orbital excitation energy should correspond to a larger *g*-factor in InAs, therefore we assign $|g_L|$ = 9.2±0.1 and $|g_R|$ = 8.9±0.1. At frequencies above 10 GHz the two resonances are more than a linewidth apart, allowing us to control the left or the right qubit separately.[8]

Coherent control over spin-orbit states is demonstrated in a time-resolved measurement of Rabi oscillations,[2,18,20] explained in Figs. 2a and 2b. Periodic square pulses shift the relative positions of the energy levels in the two dots between spin blockade (SB) and Coulomb blockade (CB). First, the double dot is initialized in a parallel state by idling in SB. This is followed by a shift to CB from which electrons cannot escape. While in CB, a resonant microwave burst is applied for a time $\tau_{burst}$ to induce qubit rotation. Finally, the double dot is brought back into SB for readout. At the readout stage the probability of the left electron to tunnel out is proportional to the singlet component of the (1,1) state. This cycle is repeated continuously.

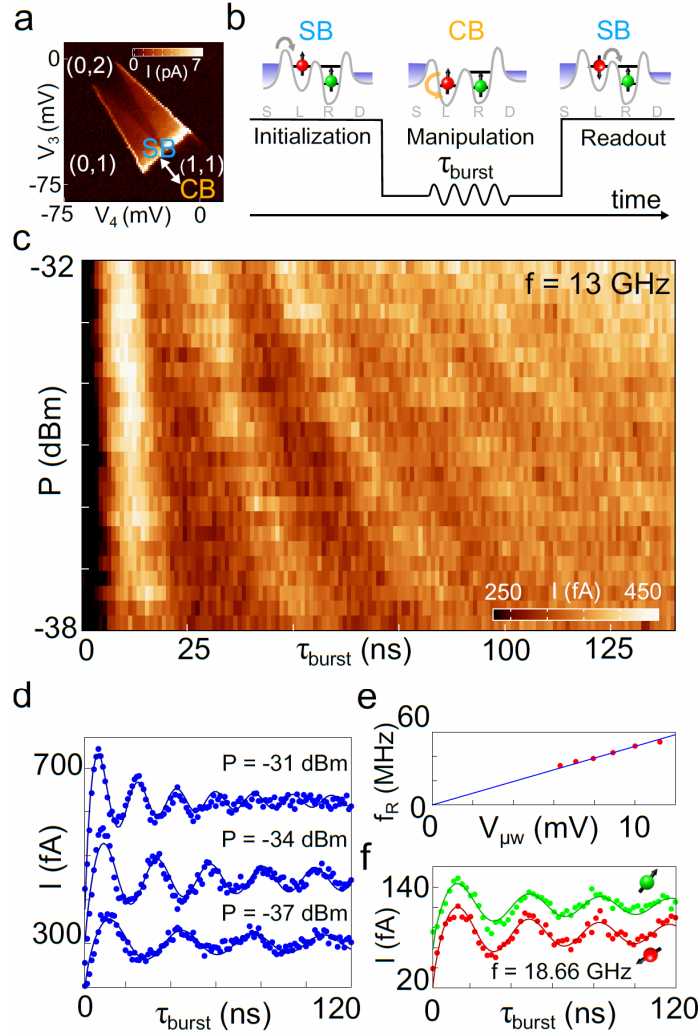

**Fig.2. Rabi oscillations. a**, charge stability diagram obtained by sweeping voltages $V_3$ and $V_4$ on gates 3 and 4. **b**, measurement cycle with diagrams showing electrochemical potentials of the source (S), drain (D), left dot (L) and right dot (R) for each stage. **c**, Rabi oscillations for a range of microwave power at $f$ = 13 GHz, $B$ = 102 mT. **d**, Rabi oscillations at $f$ = 13 GHz with fits to $a \cdot cos(f_R \cdot \tau_{burst}+\varphi)/\tau^d+b$ ($d$ = 0.8 for top trace and $d$ = 0.5 for the two bottom traces.)[21] Rabi frequencies are 58±2, 43±2 and 32±2 MHz (top to bottom). Linear slopes of 2 fA/ns, 1 fA/ns and 0.3 fA/ns (top to bottom) are subtracted to flatten the average. They are attributed to photon-assisted tunnelling. Traces are offset vertically for clarity. **e**, dependence of $f_R$ on driving amplitude $V_{\mu w} = 2(P \cdot 50\,\Omega)^{0.5}$ with a linear fit. **f**, Rabi oscillations with separated addressing of the left and right qubit at $f$ = 18.66 GHz and $B$ = 144 mT (red) and 149 mT (green) with $f_R$ = 29±2 MHz fit to the expression used in **d**.

The singlet component in the final state is measured as the d.c. current. The current oscillates as $\tau_{burst}$ is varied reflecting Rabi oscillations of the driven qubit (Fig. 2c). Rabi oscillations are observed for driving frequencies in the range $f \approx 9 - 19$ GHz. Rabi oscillations are not observed at lower frequencies (and lower magnetic fields) because the effective spin-orbit field $B_{SO} < B_N$, such that nuclear fluctuations average out the coherent qubit dynamics. We note that the observation of incoherent EDSR (Fig. 1c) requires a much smaller $B_{SO}$, because even qubit rotations with a random phase contribute to extra current near resonance.

Our highest Rabi frequency is $f_R = 58\pm2$ MHz (Fig. 2d), achieved at $f = 13$ GHz. The field $B_{SO}$ is expected to grow with $B$,[16] however at higher driving frequencies the Rabi frequency is limited by the maximum microwave source power and by the reduced transmission of the microwave circuit. With the strongest driving the amplitude of the orbital oscillation is estimated to reach 1 nm. The qubit state is flipped in ~110 microwave periods, and thus rotated by ~1.6° per cycle of the orbital motion.

We can resolve up to 5 Rabi oscillation periods. The damping of the oscillations at $P <$ -32 dBm is consistent with a $\sim(\tau_{burst})^{-0.5}$ decay envelope observed previously for rotations of a single spin interacting with a slow nuclear bath.[21] We have verified that $T_1$ relaxation does not limit coherent evolution on timescales up to 1 μs (see Supplementary Information section 3). The qubit manipulation fidelity is 48 ± 2% estimated by comparing the values of $B_{SO}$ and $B_N$ (see Supplementary Information section 5.3).[18] As expected, the Rabi frequency is proportional to the square root of the microwave power $P$ applied to the gate (Fig. 2e). Absorption of microwave photons enables interdot tunnelling regardless of the qubit state. This effect likely accelerates the decay of Rabi oscillations near the highest power (Fig. 2d, upper trace).[2,18] However, the apparent photon-assisted tunnelling is substantially reduced for $P < -32$ dBm, while Rabi frequencies remain high.

In Figures 2c, 2d only the left qubit is rotated. Figure 2f shows coherent rotations of either the left or the right qubit induced at the same microwave frequency but at two different magnetic fields, which correspond to the two EDSR resonance conditions shown in Figure 1e (see Supplementary information section 4).[22]

In the Rabi experiment the qubit state is rotated only around one axis. This is not enough for full qubit operation, which ultimately requires the preparation of an arbitrary superposition of ⇑ and ⇓, known as universal control.[23-25] Such ability is demonstrated in a Ramsey experiment (Figs. 3a, 3b). Now two short bursts with a different microwave phase are applied during the manipulation stage. In the reference frame that rotates at the Larmor frequency, the qubit is initially rotated from |+z> to |-y> on the Bloch sphere by applying a $\pi/2$ rotation around the x-axis. After a delay time $\tau$ we apply a $3\pi/2$ pulse. The tunable phase of the microwave signal $\phi$ sets the axis of the second rotation ($\phi = 0$ corresponds to a rotation around x, $\phi = \pi/2$ corresponds to a rotation around y). The final z-component depends on the axis of the second rotation as well as on dephasing. The double dot current oscillates with $\phi$ revealing Ramsey fringes (Fig. 3a). The contrast of the Ramsey fringes decreases with increasing $\tau$, allowing us to determine the inhomogeneous dephasing time $T_2^* = 8\pm1$ ns (Fig. 3b).

Coherence can be extended by a Hahn echo technique, which partially cancels dephasing coming from a slowly varying nuclear magnetic field (Figs. 3c, 3d). In the echo sequence a $\pi$ pulse is applied half way between the two $\pi/2$ pulses. The contrast of the Ramsey fringes is extended to longer coherent evolution times by performing Hahn echo (Fig. 3c). The phase of the fringes can be flipped depending on whether the $\pi$ rotation is around the x-axis ($\pi_x$) or around the y-axis ($\pi_y$). Both $\pi_x$ and $\pi_y$ Hahn echo's increase the coherence time to $T_{echo} = 50\pm5$ ns (Fig. 3d).

Gate-defined spin qubits were previously only realized in lateral quantum dots in GaAs/AlGaAs two-dimensional electron gases.[9] Due to the much stronger spin-orbit interaction in InAs, the Rabi frequencies in our InAs nanowire spin-orbit qubits are more than an order of magnitude higher than in GaAs dots.[2] Dephasing times $T_2^*$ are of the same order in InAs and GaAs quantum dots.[23,26] The relatively short $T_{echo}$ found in the present work encourages a further study. A likely reason is faster nuclear spin fluctuations caused by the large nuclear spin of indium $I = 9/2$. However, charge noise and nearby paramagnetic impurities cannot be ruled out as significant dephasing sources (see Supplementary Information section 6). Nanowires offer future solutions for suppressing the effects from nuclear spins, such as nanowires with sections of nuclear spin-free silicon. The qubit can be stored in a silicon section of the nanowire, and only moved to an InAs section for manipulation using spin-orbit interaction.

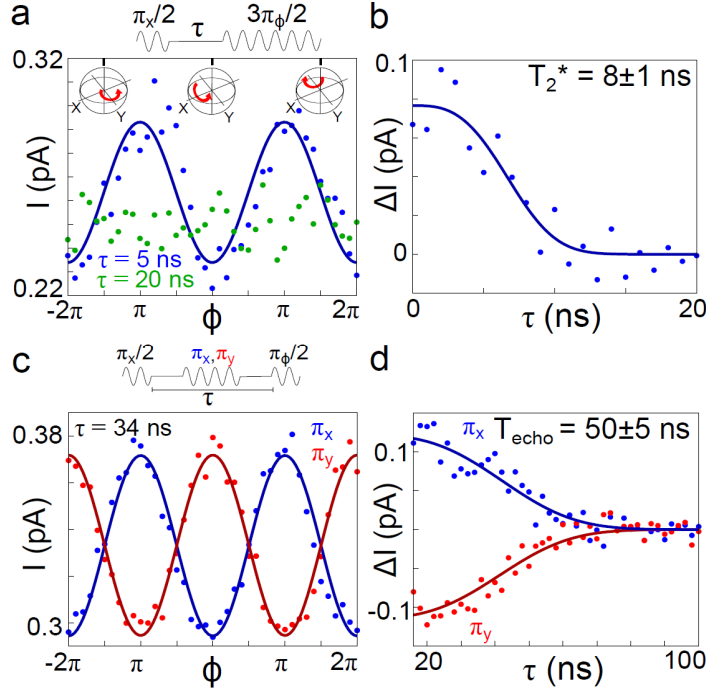

**Fig.3. Universal qubit control and coherence times. a,** Ramsey fringes schematic (top) and measurement of $I(\phi)$ for $\tau = 5$ ns and $\tau = 20$ ns. The axes of the second rotation are indicated with red arrows on the Bloch spheres for three values of $\phi$. **b,** decay of the Ramsey fringe contrast $\Delta I = I(\phi=\pi) - I(\phi=0)$ fit to $exp(-(\tau/T_2^*)^2)$. **c,** Hahn echo sequence (top) extends fringe contrast beyond $\tau = 34$ ns. Fringes for two orthogonal phases of the $\pi$ pulse are out of phase. **d,** decay of the fringe contrast obtained for the two Hahn echo sequences is used to extract $T_{echo}$ from a fit to $exp(-(\tau/T_{echo})^3)$. A fit to $exp(-(\tau/T_{echo})^4)$ gives a similar value of $T_{echo}$. In this figure the duration of a $\pi$ pulse is 14 ns using $P = -35$ dBm, $f = 13$ GHz, $B = 102$ mT.

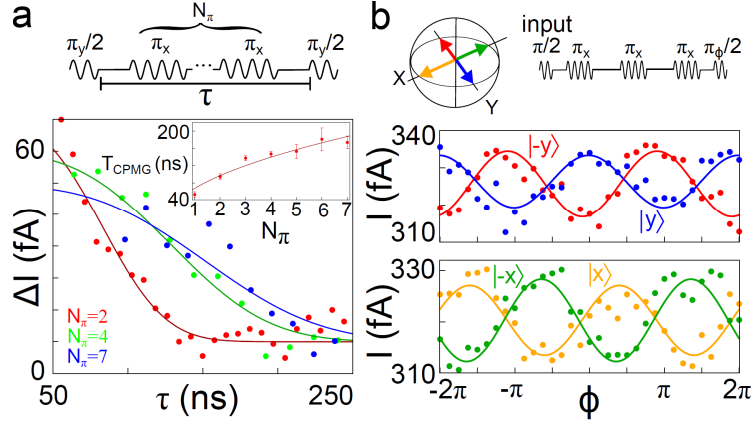

**Fig.4. Dynamical decoupling. a,** decay of the contrast of the Ramsey fringes for CPMG sequences (top) with an increasing number of π pulses $N_\pi$. Solid lines are fits to $exp(-(\tau/T_{CPMG})^3)$. Inset: the coherence times $T_{CPMG}$ vs. $N_\pi$ are fit to $(N_\pi)^d$ with $d = 0.53\pm0.1$. Error bars are standard deviations of $\Delta I (\tau)$ fits. **b,** Ramsey fringes for four different phases of the initial π/2 pulse obtained for a $N_\pi = 3$ CPMG sequence (shown above the panel) with $\tau$ = 150 ns. The input states are indicated with arrows on the Bloch sphere. In this figure the duration of a π pulse is 8 ns at $P$ = -32 dBm, $f$ = 13 GHz, $B$ = 102 mT.

Already in the present qubit longer coherence times are achieved by Carr-Purcell-Meiboom-Gill (CPMG) dynamical decoupling pulse sequences (Fig. 4a).[27,28] Now a single echo π pulse is replaced with an array of equidistant π pulses, each of which refocuses the qubit state. The total time of coherent evolution grows as the number of π pulses is increased (Fig. 4a (inset)). Importantly, arbitrary prepared qubit state in the x-y plane is preserved during the decoupling sequence. This is verified in Figure 4b which shows that the phase of the initial π/2 pulse determines the phase of the Ramsey fringes. Similar evaluation was carried out for CPMG sequences up to seven π pulses. In the future more efficient dynamical decoupling can be achieved using nuclear spin state preparation[27,29] in combination with faster π pulses or adiabatic pulse techniques.[30]

## *Methods Summary*

Devices are fabricated on undoped Si substrates. Instead of a global back gate, two wide gates B1 and B2 are located underneath the nanowire contacts (Fig. 1a). They are set to constant positive voltages to enhance conductance through the nanowire. The wide gates are covered by a 50 nm layer of Si$_3$N$_4$ dielectric, on top of this layer narrow gates and another 25 nm layer of Si$_3$N$_4$ are deposited. InAs nanowires with diameters between 50-80 nm are grown nearly free of stacking faults using metal-organic vapour phase epitaxy (MOVPE). The wires have the wurtzite crystal symmetry with the c-axis along the long nanowire axis. Nanowires are transferred in air from the mother chip to the device substrates which already contain Ti/Au gates. Selected wires are contacted with ohmic Ti/Al electrodes, during the same step contacts are made to the gates. Measurements are performed in a He$^3$ refrigerator at $T$ = 300 mK. Magnetic field is applied in the plane of the substrate at an angle of 45°±5° with respect to the nanowire. High frequency pulses are created using two arbitrary waveform generators (1 gigasample/second) and a 20 GHz / 23 dBm microwave vector source. Pulses are delivered to the sample via silver-plated CuNi coaxial lines with 36 dBm of attenuators, followed by coplanar striplines printed on the sample holder. Square pulses are applied synchronously to gates 2 and 4. Microwave bursts are applied to gate 4. A measurement cycle lasts 2 µs in Figure 2f. In the rest of the paper a cycle lasts 600 ns and each data point is averaged over 5-40 million cycles. The pulse period should remain less than 2 µs in order to detect the double dot current, limited by the noise floor of the d.c. current amplifier.

**Acknowledgements** We thank K. Nowack, R. Schouten, M. Laforest, K. Zuo, M. Hocevar, R. Algra, J. van Tilburg, M. Scheffler, G. de Lange, V. Dobrovitski, J. Danon, R. Hanson, R. Liu and L. Vandersypen for their help. This work has been supported by NWO/FOM the Netherlands Organization for Scientific Research, an ERC-Advanced Grant, and through the DARPA program QUEST.